# Advanced holeburning techniques for determination of hyperfine transition properties in inhomogeneously broadened solids applied to $Pr^{3+}:Y_2SiO_5$

Mattias Nilsson, Lars Rippe, Stefan Kröll
*Department of Physics, Lund Institute of Technology, P.O. Box 118, S-22100 Lund, Sweden*

Robert Klieber, Dieter Suter
*Fachbereich Physik, Universität Dortmund, 44221 Dortmund, Germany*

**A sequence of optical holeburning pulses is used to isolate transitions between hyperfine levels, which are initially buried within an inhomogeneously broadened absorption line. Using this technique selected transitions can be studied with no background absorption on other transitions. This makes it possible to directly study properties of the hyperfine transitions, e.g. transition strengths, and gives access to information that is difficult to obtain in standard holeburning spectroscopy, such as the ordering of hyperfine levels. The techniques introduced are applicable to absorbers in a solid with long-lived sublevels in the ground state and where the homogeneous linewidth and sublevel separations are smaller than the inhomogeneous broadening of the optical transition. In particular, this includes rare-earth ions doped into inorganic crystals and in the present work the techniques are used for spectroscopy of $Pr^{3+}$ in $Y_2SiO_5$. New information on the hyperfine structure and relative transition strengths of the $^3H_4$ - $^1D_2$ hyperfine transitions in $Pr^{3+}:Y_2SiO_5$ has been obtained from frequency resolved absorption measurements, in combination with coherent and incoherent driving of the transitions.**

## I. INTRODUCTION

Optical holeburning is a powerful tool for spectroscopy of crystalline and amorphous materials, making it possible to overcome the resolution limit imposed by inhomogeneous broadening due to strains and imperfections. Effects that can be studied include dephasing processes, spectral diffusion, hyperfine and superhyperfine interactions and Stark and Zeeman splittings[1]. The field of holeburning spectroscopy opened up with the introduction of frequency stabilised tuneable dye lasers and can continue to develop as lasers with improved stability, as well as optical modulators enabling fast frequency and amplitude modulation, become available. In this paper we describe how the technique of persistent spectral holeburning can be extended to involve a complex sequence of burning pulses at several frequencies, which tailors the inhomogeneous absorption profile to uncover information that is inaccessible in a traditional holeburning spectrum.

In addition to being a spectroscopic technique, spectral holeburning has also been the basis for numerous proposed applications, where persistent spectral holes constitute a memory mechanism that can be used for processing or storage of information. The applications include frequency-domain data storage[2], time-domain data storage[3,4], spectral-spatial correlation[5], temporal pattern recognition[6], optical data processing[7,8], radio frequency spectrum analysis[9] and the use of spectral holes as frequency references[10,11].

Rare-earth ions doped into crystalline hosts are particularly suited for study using non-linear spectroscopic methods such as holeburning[12], and for use in spectral holeburning applications. This is because, at liquid helium temperatures, the homogeneous widths of the absorption lines are many orders of magnitude smaller than the inhomogeneous broadening of the transition. This means that a laser with sufficiently narrow linewidth will only interact with a small subset of the absorbing ions. The number of addressable frequency channels, given by the ratio $\Gamma_{\text{inhom}}/\Gamma_{\text{hom}}$, can be as large as $10^7$-$10^8$ (Refs. 13,14). The narrow homogeneous linewidth corresponds to a long optical phase memory, $T_2$, according to $T_2 = (\pi\Gamma_{\text{hom}})^{-1}$, and consequently materials of this type are used in studies of optical coherent transient effects and in many applications utilising coherent transients[2,4,6,9,15-17].

Recently, rare-earth-ion-doped materials have been used in quantum optical experiments, such as in demonstrations of slow light based on electromagnetically induced transparency (EIT)[18], for the demonstration of single-photon self-interference[19] and in experiments aimed at using the dopant ions as quantum computer hardware[20,21]. Several rare-earth-ion-doped materials exhibit long-lived hyperfine or Zeeman levels and in the applications mentioned above the spin transitions and the properties of the hyperfine levels are highly relevant. The lifetime of these hyperfine levels is often many orders of magnitude longer than the optical lifetime, e.g. a few minutes in $Pr^{3+}$:$Y_2SiO_5$ (Ref. 22) and several days in $Eu^{3+}$:$Y_2SiO_5$ (Ref. 23).

The work presented here is part of a project aimed at preparing rare-earth-ions in inorganic crystals for use as quantum computer hardware[20,24,25]. As the idea is to use hyperfine levels in the optical ground state as qubit states, the ability to access and manipulate the ions on specific hyperfine transitions, as is demonstrated in the present work, is of great importance. Related results, demonstrating robust qubit manipulation and qubit distillation in $Pr^{3+}$:$Y_2SiO_5$, will be presented separately[26]. The present work relates the use of the developed techniques for spectroscopy.

Section A contains a discussion of the principles of spectral holeburning and its extensions, including the preparation of non-absorbing spectral regions and isolated spectrally narrow structures. Some comments are made on the design of holeburning sequences. Section B concerns an explicit example of a pumping sequence that prepares the inhomogeneous absorption profile of the $^3H_4 - \,^1D_2$ transition in $Pr^{3+}$:$Y_2SiO_5$ to reveal a group of ions absorbing on specific hyperfine transitions. In section C the experimental set-up is described. The results from absorption measurements, which directly yield the relative strengths of the transitions between different hyperfine levels as well as level separations and ordering, are given in section D. In section E is shown how coherent driving of the different transitions yields additional information, such as absolute Rabi frequencies. The experimental results in sectins D and E are compared with previous spectroscopic work in $Pr^{3+}$:$Y_2SiO_5$ (Refs. 22,27-29).

## II. HOLEBURNING TECHNIQUES

### A. Conventional holeburning

In spectral holeburning measurements, a laser selectively excites a narrow portion of an inhomogeneous absorption line. The absorption is bleached as absorbers are removed from the ground state and this can be seen by monitoring transmission, fluorescence or light diffracted off the spectral hole[1,30], while scanning the frequency through the originally excited frequency. The spectral resolution is determined by the linewidth of the laser and by using frequency stabilised lasers very narrow spectral structures can be studied directly in the frequency domain[31,32]. If all excited absorbers relax back to the ground state, i.e. in a closed two-level system, the method is a type of saturation spectroscopy and the hole will decay with the lifetime of the excited state. Persistent spectral holeburning occurs if some absorbers (or more generally, some absorber+host systems) are transferred to other states, with lifetimes considerably longer than that of the excited state. There exist several mechanisms for persistent holeburning, e.g. photoionisation and photochemical reactions, which was used in the first demonstrations of the phenomenon[33,34]. The mechanism that will be considered here is due to population redistribution between long-lived sublevels in the electronic ground state. In rare-earth doped solids these sublevels can be due to e.g. hyperfine, superhyperfine or Zeeman interactions[12] and in the following the term "hyperfine levels" will be used. The experiments described in this work utilises holeburning due to optical pumping[35,36] of the long-lived hyperfine levels in a rare-earth-ion-doped inorganic crystal, as was first demonstrated in Ref. 37.

In the general discussion in this section we assume absorbers with long-lived ground state hyperfine levels with energy separations $\delta_1^g$, $\delta_2^g$ etc (starting from the lowest state) and an optically excited state with hyperfine splittings $\delta_1^e$, $\delta_2^e$ etc. Since the inhomogeneous broadening is usually large compared to the hyperfine splittings, different subsets of absorbers will have different transition frequencies between ground and excited state levels. Any laser frequency inside the inhomogeneous line will be resonant with all possible optical transitions, but for different subsets of absorbers. It can be assumed that for a sufficiently long or intense pulse even weak transitions can be saturated. It is also often assumed that the absorbers relax through several complex routes and have some probability of ending up in any of the ground state hyperfine levels. Thus, by applying a burning pulse much longer than the excited state lifetime, or several pulses separated by more than the lifetime, most of the atoms absorbing at a particular frequency can be pumped to other ground state hyperfine levels. In addition to the main hole at the burning frequency, this creates a pattern of sideholes due to transitions from the ground state levels with depleted population to each of the excited state levels. Antiholes, i.e. increased absorption, will occur at frequencies corresponding to transitions from the ground state levels to where the absorbers have been pumped. If the sideholes and antiholes can be resolved, their positions directly give the hyperfine level separations in the ground and excited states. In principle, the relative amplitudes of the holes can give information about the strength of the transitions between different hyperfine levels, but this requires information or assumptions about branching ratios and about the relaxation rates between the ground state hyperfine levels. Fig. 1 illustrates a simple case of hyperfine holeburning.

### B. Breaking the equivalence between the pumped hyperfine transitions

An important feature of the spectral holeburning techniques described in this work is that they employ optical pumping at a selected combination of frequencies within the inhomogeneous absorption profile. As an example, consider the advantages of using pumping light at two frequencies: In a holeburning spectrum created using a single burn frequency, the spectral positions of sideholes and antiholes are symmetric and it is not possible to determine the ordering of hyperfine levels (corresponding to the signs of hyperfine coupling constants) from this. However, by burning at two frequencies, with a separation equal to one of the separations between the ground state hyperfine levels, some absorbers can be pumped away from two of the hyperfine levels. This breaks the symmetry, as anti-holes at specific positions will increase in size when the population in the corresponding hyperfine levels increase. Comparing the resulting holeburning spectrum with hole patterns created using a single burn frequency directly yields the ordering of the ground state levels.

In general, the sequences described below involve burning at three or more frequencies chosen such that, for a subset of absorbers at a specific frequency within the inhomogeneous line, these frequencies correspond to excitation from each of the ground state hyperfine levels. Absorbers not belonging to this subset, although resonant with at least one of the burn frequencies, are pumped to levels where they are not affected by subsequent laser pulses. Similar techniques have been used implicitly in experiments concerning EIT and slow light[18,38], where the process involves excitation from two of the ground state hyperfine levels and an additional field is used to re-pump absorbers from the third level. Unless similar schemes are followed, any experiment involving excitation on specific hyperfine transitions hidden by a large inhomogeneous optical broadening will be complicated by the fact that each laser field will also excite other transitions for subsets of absorbers that have the appropriate transition frequencies (c.f. Fig. 1).

### C. Creating a spectral pit

An important part of the techniques used in this work is the ability to completely remove all absorbers from an interval within an inhomogeneous absorption line of (Fig. 2a). This is done by repeatedly scanning the frequency of the burning field across the spectral region, thus burning a wide hole or "pit" in the absorption profile. Within the pit, there is essentially no absorption or fluorescence, so if a set of absorbers is prepared at a specific frequency within the pit, measurements can be made on zero background.

The amount of absorbers within a spectral interval can be reduced to a fraction of a percent by repeated or continuous pumping. The minimum obtainable absorption is determined by the spectral purity of the light and by the relaxation rate of the pumped levels compared to the pumping rate. The spectral purity is important because any energy outside the pumped spectral region can be absorbed. Light outside the pit may also excite other transitions, thus pumping absorbers back into the emptied region. Relaxation may be limiting if the lifetime of the hyperfine levels is not several orders of magnitude longer than that of the optically excited state used in the pumping process, since it wil be necessary to cycle the optical transition many times before the pit is empty.

One question that may be considered is how wide a spectral interval can be emptied. This is relevant because it determines whether it is possible to study several transitions at different frequencies inside a single pit, as is demonstrated below, and because it determines the maximum bandwidth that can be used in an experiment without affecting "spectator absorbers" outside the pit. It is clear that the emptied region can never be broader than the total energy separation between the ground state hyperfine levels ($\delta_{tot}^g = \delta_1^g + \delta_2^g + ...$), since this is the maximum frequency distance the absorption due to a particular absorber can be shifted, by moving the absorber between hyperfine levels. Put differently, if a spectral interval smaller than $\delta_{tot}^g$ is repeatedly excited, the absorbers may eventually decay to a hyperfine level where they only absorb outside this interval. This will be the case e.g. if all excited state hyperfine splittings are larger than the total splitting of the ground state ($\delta_{tot}^g < \delta_1^e, \delta_2^e, ...$). Another case is when the total splitting of the excited state, $\delta_{tot}^e$, is smaller than the total splitting in the ground state. In this case, a pit with a width approaching ($\delta_{tot}^g - \delta_{tot}^e$) can be created. The analysis will be more complex if several close-laying pits are created.

### D. Creating an isolated peak

Another key technique is the method of creating a spectrally narrow peak of absorbers in a pit, i.e. in region that has been emptied of all other absorbers (Fig. 2b). Fluorescence measurements can then be made on zero background and as long as the peak is narrower than the splitting between hyperfine levels, different transitions will appear as absorbing peaks at different frequencies, which can be directly compared. Perhaps more importantly, as was demonstrated by Pryde et al.[39], it is possible to apply "hard" optical pulses, i.e. pulses with a specific pulse area, to a narrow and spectrally isolated absorptive feature. This makes it possible to use (multiple pulse) coherent transient techniques frequently used in NMR[40], which can provide a wealth of structural and dynamical information.

Two methods for creating a spectrally isolated peak have appeared in literature. The first method relies on the ability to excite absorbers in the surrounding interval, leaving a central group of absorbers in the ground state. Some of the excited absorbers will relax to other states and if the procedure is repeated many times an isolated absorbing peak will remain. Pryde et al. achieved this by using phase modulated pulses to achieve spectrally selective excitation[39,41]. The principle is similar to that demonstrated by Kikas and Sildos[42], who used one laser to excite a broad spectral interval and a second laser to deexcite a narrow region in the centre of this interval. This method of generation of a narrow absorptive feature has also been explored by de Sèze et al., using chirped pulses in the regime of rapid adiabatic passage[43]. The minimum attainable width of a peak created by repeated burning on the surrounding spectral interval is determined by laser jitter during the pulse sequence. The method does not select atoms in a specific hyperfine state or absorbing on a specific transition.

In the other method, introduced in Ref. 20, a spectral interval is first completely emptied of absorbers using repeated pulses chirped over the frequency interval. The frequency is then shifted by an amount equal to one of the ground state hyperfine level separations, $\delta_i^g$, and absorbers within a narrow frequency interval are excited. Some of the absorbers will decay into the previously emptied hyperfine level and will now absorb in the centre of the emptied spectral region, on transitions going from a single hyperfine level. If the relevant relaxation branching ratio is sufficiently large, this "burning back" into the pit needs only be repeated one or a few times in order to create a significant peak. Thus the width of the peak, created using this method, is only limited by the short-term frequency stability of the laser source. The width of the created peak may also be increased by inhomogeneous broadening of the hyperfine level separation. In this case, it is possible to optically probe inside an inhomogeneously broadened spin transition, directly in the frequency domain, or to apply additional burning pulses to narrow the peak. As hyperfine level separations are usually isotope dependant, an absorbing peak created using this method will be due to only one isotope, even though all isotope shifts are usually buried by the optical inhomogeneous broadening.

### E. Holeburning sequences

The techniques described above are combined to prepare groups of atoms, absorbing on specific transitions and on zero background, inside the inhomogeneously broadened absorption line. These atoms can then be used for spectroscopic studies, as below, or for experiments within quantum optics or quantum information processing[20,21,26].

A typical burn sequence will start with a series of pulses that, to the extent possible, removes subsets of atoms that absorb on other hyperfine transitions than those selected, from the spectral region that will be used. Then a pit (pits) are burned around the frequency (frequencies) where the selected subset of atoms are to absorb on specific transitions. In the process of emptying the pits, these absorbers are moved to an auxiliary state. Finally, pulses resonant with transitions going from this auxiliary state pump the selected subset of absorbers back into the emptied spectral region. Additional tailoring pulses may be needed in order to completely spectrally isolate the selected absorbers. Figure 3 illustrates a basic holeburning sequence, preparing an isolated set of absorbers in a specific ground state sublevel.

Considering that a single frequency burn pulse, applied to a sample with three sublevels in both ground and excited state, changes the absorption at 48 other frequencies (6 sideholes and 42 antiholes), the frequencies at which to burn must be chosen with some care. In general there will be restrictions on what transitions should be used, how large pits can be created, etc, imposed by coincidences between transitions for different subsets of absorbers and by the laser linewidth. Fortunately, with knowledge about the hyperfine level separations obtained using standard holeburning, it is possible to simulate the burning kinetics and to predict where the different subsets of atoms will absorb after the application of a sequence of burning pulses. Such simulations were used in developing the burn sequences described below.

## III. HOLEBURNING SEQUENCES FOR ISOLATING TRANSITIONS IN $Pr^{3+}:Y_2SiO_5$

We now turn to the burn sequences used to reveal specific hyperfine transitions of the $^3H_4 - {}^1D_2$ transition of $Pr^{3+}$ doped $Y_2SiO_5$. The Pr ions can occur in two crystallographically inequivalent sites and here we only consider ions in site I, absorbing at 606 nm. Praseodymium has a single naturally occurring isotope ($^{141}Pr$) with a nuclear spin I=5/2. Each electronic state is split into three doubly degenerate states by second order hyperfine interaction and electric quadrupole interaction[12]. The hyperfine level separations, shown in Fig. 4, are between 4.6 and 27.5 MHz and the states are conventionally labelled ±1/2, ±3/2 and ±5/2. However, the nuclear wavefunctions are mixed due to the low site symmetry and the states are not strictly spin eigenstates[22,28]. As a result, transitions between any pair of hyperfine levels are allowed to some degree. At cryogenic temperatures (~4 K), relaxation between ground state hyperfine levels, due to phonon assisted processes or interaction with nuclear magnetic moments in the host, occur at a rate of about $(100\ s)^{-1}$. The lifetime of the excited state, $T_1$, is 164 µs and the time required to saturate the transition with laser radiation is typically 1-100 µs, which means that thousands of burning cycles can be performed before hyperfine relaxation becomes significant.

In Appendix A we give, as an explicit example, the details for a sequence of burning pulses that prepares a group of ions in the ±1/2 level in the ground state, ±1/2(g), such that their absorption to each of the excited state hyperfine levels occurs within an otherwise empty spectral region. A frequency $\nu_0$ is selected within the inhomogeneous absorption line and burning is mainly done around the frequencies $\nu_0$, ($\nu_0$+10.2 MHz) and ($\nu_0$+27.5 MHz), which for the selected subset of ions correspond to transitions 1/2(g)-1/2(e), 3/2(g)-1/2(e) and 5/2(g)-1/2(e), respectively. The first pulses remove other subsets of ions, to the extent possible. Then a wide pit is burned around frequencies $\nu_0$ and ($\nu_0$+10.2 MHz), which leaves the selected subset of ions in the ±5/2 hyperfine level of the ground state. Finally, ions are excited on the ±5/2(g)-±5/2(e) transition and relax back into the ±1/2 state, where they form a narrow absorbing feature. The sequence consists of more than 500 pulses, with durations between 20 and 200 µs, spread out during approximately 0.5 s. Between experiments the hyperfine level populations can be scrambled by repeated burning on a wide interval, which creates an equilibrium-like state that is taken as a starting point for repeated experiments.

Figs. 5b-d show the relative number of $Pr^{3+}$ ions absorbing on each of the nine possible transitions at each optical frequency, after the burning sequence in Appendix A has been applied. The figures are the result of a simulation of the burning dynamics assuming initial equilibrium, with equally many ions in each ground state hyperfine level. It has been assumed that all transitions can be saturated and that relaxation to each of the ground state levels is equally probable. Even if this is not strictly true, the simulation correctly shows the positions of the created absorbing peaks and that the surrounding spectral interval may be completely emptied with sufficient burning.

Fig. 5a shows the results of applying the burn sequence experimentally, where readout is performed by monitoring the transmission of a frequency chirped pulse. The group of ions prepared in the ±1/2(g) state will give rise to peaks of absorption at frequencies corresponding to the transitions to each of the excited state hyperfine levels. The relative transition strengths of the transitions can be directly seen and the properties of the transitions and hyperfine levels can be further studied by applying additional pulses to the peaks.

## IV. EXPERIMENTAL SET-UP

The experimental set-up is shown in Fig. 6. A $Nd:YVO_4$ laser (Coherent Verdi) pumped a ring dye laser (Coherent 699-21) using Rhodamine 6G to give ~300 mW output power at $\lambda$=606 nm. The dye laser was actively stabilised, using an intra-cavity electro-optic modulator (Gsänger PM25), and was frequency locked to a stable reference cavity, which reduced frequency jitter to <30 kHz. The laser light was modulated by an acousto-optic modulator (AOM) (A.A.Opto-Electronics), with a bandwidth of 100 MHz, used in a double-pass configuration in order to increase the bandwidth and to eliminate beam movement accompanying frequency shifts. The AOM was driven by a 1 GS/s arbitrary waveform generator (Tektronix AWG520), which allowed direct control of pulse amplitude, phase and frequency. The waveform generator was computer controlled, which facilitated programming of complex pulse sequences. A single mode optical fibre was used as a mode cleaner after which a beam sampler (Thorlabs BSF10-A1) picked off 5% of the light for use as a reference beam. The burn/probe beam passed a $\lambda$/2-plate and was then focused onto the sample using an f=300 mm lens. The diameter ($I=I_0/e^2$) of the focus was ~100 µm throughout the sample. The optical power at the sample was ~20 mW, which gave Rabi-frequencies of around $\Omega \approx 1$ MHz when the transitions were coherently driven.

The sample was a 0.5 mm thick $Y_2SiO_5$ crystal with a $Pr^{3+}$ concentration of 0.05 at.%. The crystal was held at a temperature of ~3.5 K. The light propagated along the b axis and the polarisation was linear and was rotated, using the $\lambda/2$-plate, to obtain maximum absorption. This polarisation was taken to be along a direction where the E-field projection is equal onto both possible orientations of the $Pr^{3+}$ site I.

The focus in the sample was imaged onto a 50 $\mu$m pin-hole, in order to study only ions in the centre of the beam, where the intensity variations were small (<20%). The transmitted beam and the reference beam were detected using a matched pair of electronically amplified photo-diodes. Two AOMs were used to gate the detection, blocking the burning pulses.

Read-out of spectral structures, prepared by holeburning, was performed by scanning the light frequency and recording the intensities of the transmitted beam and the reference beam. The signals were then divided in order to remove shared noise, such as laser amplitude noise. The probe pulses were short (~100 $\mu$s) and a factor $10^3$-$10^4$ weaker than the burning pulses, in order to minimise excitation during read-out. The frequency scan rate, r, of the read-out was typically between 0.01 MHz/$\mu$s and 1 MHz/$\mu$s. The spectral structures that were prepared, e.g. absorbing peaks, had features that were less than 100 kHz wide, determined by laser jitter and inhomogeneous broadening of the hyperfine transitions. However, the read-out resolution was limited by the Fourier width set by the read-out scan rate, $\Delta \upsilon \propto \sqrt{r}$. All data were obtained in single-shot measurements, unless otherwise stated.

## V. ABSORPTION MEASUREMENTS AND INCOHERENCT PROBING OF TRANSITIONS

Using three different holeburning sequences of the type described above, a peak of absorbing ions was isolated in the $\pm 1/2$, $\pm 3/2$ or $\pm 5/2$ hyperfine level of the ground ($^3H_4$) state, such that the absorption of the ensemble to each of the three upper state ($^1D_2$) hyperfine levels could be seen on a zero background (Fig. 7). As the same ensemble of ions was absorbing on three different transitions in each measurement, the relative strength of these transitions could be directly obtained by comparing the size of the absorbing peaks, independent of what number (density) of ions had been prepared. This was used to calculate a complete matrix describing the relative oscillator strengths for optical transitions between the ground and excited state hyperfine manifolds (Table 1). Although there was no direct connection between the measured absorption from different ground state hyperfine levels, since the number of ions prepared may have been different, the sum of the relative strengths of transitions going to each excited state was found to be approximately one, as should be expected. The results in Table 1 were obtained by constraining each row and column of the matrix to sum to one and fitting the entries to several series of absorption data.

As some of the transitions were almost too weak to be observed directly and because additional, unrelated, peaks were sometimes present, experiments were made in order to verify the assignment of the absorbing peaks to specific transitions. This illustrative experiment also gave additional information about the branching ratios for relaxation from a particular excited state hyperfine level to each of the ground state hyperfine levels. Figs. 8b-d show the result of burning on, in turn, the frequencies corresponding to the ±3/2 - ±1/2, ±3/2 - ±3/2 and ±3/2 - ±5/2 transitions. If a series of burning pulses were applied at e.g. the ±3/2 - ±1/2 transition, the ions were pumped away, via the ±1/2(e) level, and the peaks corresponding to transitions ±3/2 - ±3/2 and ±3/2 - ±5/2 disappeared, in addition to the peak at the burning frequency, which confirmed that the absorption was due to the same ensemble of ions. Similar experiments were done on the transitions from the ±1/2(g) (c.f. Fig. 7a) and ±5/2(g) (c.f. Fig. 7c) levels of the ground state. This worked even when the absorption on the pumped transition was too weak to observe directly. In Fig. 8 it can be seen that a significant fraction of the ions excited to the ±1/2 or ±3/2 levels of the excited state relax to ±1/2 in the ground state, where their absorption on transitions ±1/2 - ±3/2 and ±1/2 - ±5/2 can be monitored in the same pit. However, when the pumping is performed via the ±5/2(e) state (Fig. 8d), almost all absorbing ions disappear from the pit, which means that the branching ratio for relaxation to ±5/2 in the ground state is large. Branching ratios found by such measurements largely agreed with the oscillator strengths found for the optical transitions. This could be an indication that a significant fraction of the excited ions relaxes by spontaneous emission directly to the ground state or that spin selection rules are conserved also in more complex relaxation routes.

The results shown in Fig. 7 directly yield the separation between the three doubly degenerate hyperfine levels of the excited state, given by the separation between peaks corresponding to excitation to each of the states. This is of some interest because contradictory energy level ordering within the hyperfine manifolds in $Pr^{3+}:Y_2SiO_5$ have appeared in literature[22,27-29] and because commonly used techniques, such as optically detected nuclear magnetic resonance, only determines absolute values of the hyperfine splitting. Our results show that the separation between the lowest and the middle hyperfine level is 4.58±0.02 MHz and that the separation between the middle and the highest level is 4.86±0.02 MHz. This confirms the conclusions about level ordering drawn by Holliday et al.[22] by comparing the relative size of holeburning features in experiment and simulations. Our result corresponds to a situation where the pseudo-quadrupole coupling constants[12,44] D and E have negative signs in the ground state and positive signs in the excited state. This disagrees with the assumptions of Longdell et al.[28] and Ham et al.[29].

Longdell et al. inferred the directions of the Zeeman and pseudo-quadrupole tensors in $Pr^{3+}:Y_2SiO_5$ by measuring the hyperfine splittings while rotating the direction of a weak magnetic field[28]. From this the relative strengths of the optical transitions was calculated, as the overlap of the nuclear states between the ground and optically excited hyperfine manifold. To first order this gives the relative size of the transition dipole moments, µ, and the optical oscillator strength is proportional to the square of this. The observed transition strengths (Table 1) agree with their results as to the relative significance of the transitions, e.g. the conclusion that the transitions ±1/2 - ±3/2 and ±3/2 - ±1/2 are the strongest off-diagonal transitions. However, their calculations imply that the oscillator strengths of the diagonal transitions are more than 20 times larger than that of any of the off-diagonal transitions, which was not observed in our experiments. One possible cause for the discrepancy can be that hyperfine transitions in $Pr^{3+}:Y_2SiO_5$ often involve a simultaneous change of a Pr and Y nuclear spin[45]. This could mean that at low applied magnetic fields the optical transition does not correspond to a pure Pr transition but to a combination of near degenerate Pr-Y transitions.

Fig. 9 shows a high-resolution holeburning spectrum, recorded using an unfocused beam, together with a simulation using the parameters in Table 1. For the simulation it was assumed that spin relaxation is negligible and the quadrupole coupling constants were assumed to be negative for the ground state and positive for the excited state. The rate equations were solved numerically for the duration of the burn pulse (1 s). The simulation results can been seen to agree well with the observed hole-burning spectrum, which supports the results obtained above.

## VI. RABI FREQUENCY MEASUREMENTS

By applying intense coherent laser pulses to a transition, the absorbers can be driven into an arbitrary superposition of the two levels connected by the transition. In the Bloch sphere picture this corresponds to rotating the state vector around a vector representing the interaction with the applied electric field, at a rate proportional to the amplitude of the field[46,47]. When the radiation has a fixed frequency, resonant with the transition, the absorbers will be driven completely up into the excited state and then down again into the ground state, in a process known as Rabi flopping. The rate at which the absorbers are rotated between the states is known as the Rabi frequency, $\Omega$, and is proportional to the optical transition dipole moment, $\mu$, of the transition, according to $\Omega = 2\pi E\mu/h$. The pulse area, $\theta$, is the angle through which the state vector has been rotated, given by $\theta = \Omega t$. Rabi frequency measurements are a good way of determining the strength of coupling between light and dopant ions, but may be complicated if several transitions of different strength are simultaneously driven, or if the ions are located in sites with different orientations relative to the electromagnetic field[48].

The transition strengths of the nine transitions between the ground and excited state hyperfine manifolds in $Pr^{3+}$:$Y_2SiO_5$ were determined by studying Rabi flopping through coherent driving of the transitions. Spectrally isolated peaks, due to ions absorbing on a specific transition, were prepared as described in Sections B and D. Then coherent pulses with varying pulse areas were applied to the peaks, in a series of experiments, and the effect was studied by rapidly scanning the laser frequency across the transition while monitoring the transmission. When the ions were driven more than halfway up into the excited state, i.e. for $\theta > \pi/2$, a majority of the population was in the excited state and stimulated emission could be observed instead of absorption. The absorption/emission showed an oscillatory behaviour as a function of the applied pulse area, from which the maximum Rabi frequency could be determined (Fig 10). Because the peaks have a finite width, the applied pulses were kept relatively short ($\leq 3$ μs), in order to make them spectrally flat across the peak. Otherwise, ions absorbing at slightly different frequencies within the peak would be driven at different Rabi frequencies. Similarly, ions at different spatial positions within the laser beam have different Rabi frequencies due to variations in field intensity. This was partly compensated for by the use of a pinhole on the detection side of the set-up, which imaged only the centre of the beam. However, remaining inhomogeneity in the Rabi frequency led to a rapid decay of the Rabi oscillation, as the ions were driven more than one Rabi period. The signal was also decreased by the ~20 μs delay between the applied pulse and the moment when the read-out scan passed the centre of the peak. During this delay some of the excited ion population has had time to relax back to the ground state. The maximum applied pulse area was limited to $<3\pi$, by the available laser power and the need to keep the pulses shorter than ~3 μs.

Even the weakest transitions between the $^3H_4$ and $^1D_2$ hyperfine states could be driven coherently. However, on these transitions the absorption/stimulated emission was too weak for the population difference between the ground and excited states to be studied directly. In this case, the state after the application of the driving pulse was observed indirectly by monitoring the number of ions left in the ground state, through their absorption on a stronger transition. For example, when the $\pm 5/2(g) - \pm 1/2(e)$ transition was driven, the absorption on the strong transition $\pm 5/2(g) - \pm 5/2(e)$, 9.4 MHz away, could be seen to oscillate between a maximum value and a minimum value, corresponding to maximum ($\theta = \pi$) and minimum excitation on the $\pm 5/2(g) - \pm 1/2(e)$ transition (Fig. 11). Similarly, the driving of a transition could be monitored by studying absorption/emission on a transition connecting the excited state in the driven transition to another ground state hyperfine level. Initially there would be no absorption on the monitored transition, but as ions were moved to the excited state by the applied pulse (e.g. for $\theta \approx \pi$), negative absorption, i.e. stimulated emission, could be seen on the transition.

The Rabi frequency was measured for each of the nine transitions and the results are summarised in Table 2. The Rabi frequencies ranged from <0.2 MHz to 1.8 MHz, using 20 mW optical power focused to 100 µm. From this, the transition dipole moment of the strongest transition ($\pm 5/2 - \pm 5/2$), along the polarisation of the light field, was calculated to be $\mu = 3.7 \cdot 10^{-32}$ Cm (0.011 D), corresponding to an oscillator strength of $f = 9.5 \cdot 10^{-7}$. This is slightly higher than previously reported values ($f = 3 \cdot 10^{-7}$ (Ref. 27) and $f = 7.7 \cdot 10^{-7}$ (Ref. 49)), which is reasonable since the conventional measurement techniques give the average over all transitions between the hyperfine manifolds. The relative magnitude of the square root of the entries in Table 2 can be compared to the results in Table 1 and are in reasonable agreement. The results obtained using coherent driving of the transitions suffer from a greater uncertainty, partly because they were obtained in separate experiments while in the absorption measurements the strength of three transitions were obtained in a single shot. However, the experiments described in this section provides an additional way of comparing the strength of transitions going from different ground state hyperfine levels, which was only done indirectly in the absorption measurements.

In closing, we note that coherent driving of transitions gives an additional tool when tailoring the inhomogeneously broadened absorption profile for spectroscopy or applications. Using one $\pi$-pulse to lift ions to the excited state and another to drive them down into another hyperfine level in the ground state can be a much more efficient way of moving population between states than incoherent burning[26]. Such coherent excitation and coherent Raman transfer can be used to speed up and improve the preparation of absorbers using holeburning sequences and was used in the final stages of the work presented here.

## VII. SUMMARY

We have described spectral holeburning techniques for tailoring inhomogeneous absorption lines in solids with long-lived hyperfine levels in the ground state. This includes techniques for emptying wide spectral regions inside the absorption line and within them placing spectrally isolated absorbing features, which are useful for applications and spectroscopy. A sequence of optical holeburning pulses was used to reveal transitions between specific hyperfine levels, initially buried within an inhomogeneously broadened absorption line, such that they could be studied with no background absorption on other transitions. This was illustrated by studying the properties of the hyperfine transitions of the $^3H_4$ - $^1D_2$ transition in $Pr^{3+}$:$Y_2SiO_5$ in some detail. The relative and absolute transition strengths were determined by comparing absorption strengths and by measuring the Rabi frequency as the transitions were coherently driven.

## ACKNOWLEDGMENTS

This work was supported by the ESQUIRE project within the IST-FET program of the EU, the Swedish Research Council and the Knut and Alice Wallenberg Foundation.

## APPENDIX A: PULSE SEQUENCE FOR ISOLATING TRANSITIONS FROM THE ±1/2 STATE IN $Pr^{3+}$:$Y_2SiO_5$

Here we give the details for a sequence of burning pulses that prepares a group of ions in the ±1/2 level in the ground state, ±1/2(g), such that their absorption to each of the excited state hyperfine levels occurs within an otherwise empty spectral region. The resonance frequency $\nu_0$ of the 1/2 – 1/2 transition for this subset of ions can be chosen arbitrarily anywhere inside the inhomogeneous absorption line. The frequency offsets below are given with respect to this frequency. Numbers given are those used in the experiments.

1. Burning on the intervals –3 to 3 MHz, 7.2 to 13.2 MHz ($\delta_1^g \pm 3\,\text{MHz}$) and 33.9 to 39.9 MHz ($\delta_1^g + \delta_2^g + \delta_1^e + \delta_2^e \pm 3\,\text{MHz}$). E.g. 300 excitation pulses, 50 μs long and separated by 200 μs. This removes any ions that do not have their 1/2 – 1/2 transition within 3 MHz from $\nu_0$ and that can be moved to hyperfine levels where they only absorb outside the region under preparation. The subset of ions we wish to study cannot escape outside the pumped regions.
2. Burning on the interval -3 to 12 MHz. E.g. 300 excitation of 50 μs duration, with a separation of 200 μs. This empties a pit around the frequencies where the selected subset of ions will absorb from the ±1/2 and ±3/2 levels of the ground state. These ions are pumped to level ±5/2(g) by this step.
3. Burning at 36.9 MHz ($\delta_1^g + \delta_2^g + \delta_1^e + \delta_2^e$). E.g. 10 excitation pulses of 50 μs duration and with the laser intensity attenuated by a factor 10. This burns back a peak into the pit, i.e. excites the selected subset of ions from ±5/2(g) to ±5/2(e), from where they can relax to ±1/2(g) and ±3/2(g).
4. Readout: The laser intensity is attenuated by a factor $10^3$ and the frequency scanned e.g. between –2 and 12 MHz in 500 μs.
5. Erase structure: By repeated burning on a wide interval, e.g. –100 to 100 MHz, the hyperfine level populations can be scrambled, creating an equilibrium-like state that can be used as a starting point for repeated experiments.

# Captions

*Figure 1*. Principles of spectral holeburning through optical pumping and redistribution of absorbers between sublevels in the ground state. The example assumes absorbers with two sublevels in both ground and excited states and that the splitting of the ground state is larger than that of the excited state. (a) Due to inhomogeneous broadening, different sub-sets of absorbers will have transitions between different sublevels shifted into resonance with the applied light. The figure shows the energy levels of four different subsets, all resonant with the burning laser, but on different transitions. (b) Absorption spectrum after burning at frequency $\nu_{burn}$. The position of side-holes and anti-holes directly give the separations between the sublevels.

*Figure 2*. Spectral holeburning in the inhomogeneously broadened $^7F_0$ - $^5D_0$ transition in $Eu^{3+}$:$YAlO_3$. The amount of $Eu^{3+}$ ions absorbing at a particular frequency has been monitored by recording laser induced fluorescence. (a) A 12 MHz wide spectral interval has been completely emptied of absorbing ions. At laser frequencies inside the pit, the signal is constant and less than 1% of the fluorescence signal before burning. (b) An ensemble of ions has been moved back into the pit by shifting the laser frequency 69 MHz from the centre of the pit and thus burning on an auxiliary hyperfine level in the ground state. The resolution of the spectral features in this experiment was limited by laser jitter of the order of 2 MHz.

*Figure 3*. Simplified holeburning sequence, assuming absorbers with three sublevels in the ground state and a single excited state. Energy level diagram is shown for a selected subset of absorbers at a specific position within the inhomogeneous absorption line. (a) Burning at frequencies $\nu_0$, $(\nu_0 + \delta_2^g)$ and $(\nu_0 + \delta_1^g + \delta_2^g)$ removes absorbers for which these frequencies correspond to other transitions than for the selected subset. (b) Burning at frequencies $\nu_0$ and $(\nu_0 + \delta_2^g)$ creates two wide spectral pits and collects the selected subset of absorbers in an auxiliary state. (c) Absorbers within a narrow spectral interval at $(\nu_0 + \delta_1^g + \delta_2^g)$ are excited from the auxiliary state, from where they relax to form a narrow peak, absorbing from a selected ground state sublevel.

*Figure 4*. Spectrum and relaxation data for the $^3H_4$ - $^1D_2$ transition of site I $Pr^{3+}$ ions in $Y_2SiO_5$. $T_1$, $T_2$ and $\Gamma_{hom}$ are given according to Ref. 27.

*Figure 5*. (a) The experimental result of applying the holeburning sequence detailed in Sect. B to $Pr^{3+}$:$Y_2SiO_5$. A group of ions is prepared in state ±1/2, such that they absorb to each of the excited state hyperfine levels on frequencies inside a previously emptied spectral pit. In this experiment, the spectral resolution was limited by a fast read-out chirp rate (r≈1 MHz/μs) and the peak corresponding to the transition ±1/2-±5/2 was not resolved. C.f. Fig. 6a for a close-up of the central region. (b)-(d) Relative number of ions absorbing at each optical frequency from the three ground state hyperfine levels, after applying the sequence of burning pulses in Sect. B. (b) Ions in state ±1/2, (c) ions in state ±3/2 and (d) ions in state ±5/2. Each ion can absorb at three different frequencies. Solid lines: Absorption to the ±1/2 hyperfine level in the excited state. Dashed lines: Absorption to excited state ±3/2. Dotted lines: Absorption to excited state ±5/2.

*Figure 6*. Set-up used in the experiments. See the text, Sect. IV, for details.

*Figure 7*. Ions absorbing on specific hyperfine transitions, inside the inhomogeneously broadened $^3H_4$ - $^1D_2$ transition in $Pr^{3+}$:$Y_2SiO_5$. The three curves correspond to three separate experiments. The three peaks of varying size, in each curve, are due to the same ensemble of ions absorbing on different transitions. The intensity was attenuated by a factor $10^3$ during read-out, in order not to cause significant excitation when probing the transitions. The scan rate during read-out was of the order of r≈ 0.1 MHz/μs. Single-shot measurements.

*Figure 8*. Incoherent burning on transitions going from state ±3/2. (a) A group of ions have been prepared in state ±3/2, inside a wide region emptied of other absorbing ions. Absorption on transition ±3/2-±5/2 is barely detectable. (b)-(d) In three separate experiments, a large number of burning pulses are applied at the frequencies indicated in (a). When the ions are pumped away via (b) the ±1/2 or (c) the ±3/2 hyperfine levels in the excited state, a large fraction of the ions relax to ±1/2 in the ground state and are detected after the burning. (d) When the ions are pumped away via the ±5/2 excited state, they disappear from the monitored region, indicating that they have relaxed to ±5/2 in the ground state, since all transitions from this state are resonant with higher frequencies. Moderate burning at other frequencies inside the pit, i.e. not resonant with any of the peaks, does not change the absorption spectrum. Single-shot measurements.

*Figure 9*. Solid line: Experimental holeburning spectrum in the $^3H_4$ - $^1D_2$ transition in site I $Pr^{3+}$ in $Y_2SiO_5$. The hole was created by applying a burn pulse during 1 s. The data is an average of 30 read-out scans. The spectral resolution is limited by laser jitter during burning and read-out. Dashed line: Simulated holeburning spectrum using hyperfine parameters from Table 1.

*Figure 10.* Coherent driving of the ±1/2-±3/2 transition. Pulses with a variable pulse area ($\theta = \Omega t$) was applied to the transition using 3 μs long pulses with varying intensity. The population difference between the ground (±1/2(g)) and optically excited (±3/2(e)) states, after the pulse, was monitored by recording the transmission at the resonance frequency. (b)-(e) show an ensemble of ions absorbing/emitting on the transition after the application of a pulse with (b) $\theta = 0$, (c) $\theta \approx \pi/3$, (d) $\theta \approx \pi$ and (e) $\theta \approx 2\pi$.

*Figure 11.* Coherent driving of the ±5/2-±1/2 transition. Pulses with a variable pulse area ($\theta = \Omega t$) was applied to the transition using 3 μs long pulses with varying intensity. The number of ions left in the ground state, ±5/2(g), after the pulse, was monitored by recording the absorption on the ±5/2-±5/2 transition. (b)-(d) show an ensemble of ions absorbing on ±5/2-±5/2 after the application of a pulse with (b) $\theta = 0$, (c) $\theta \approx \pi$ and (d) $\theta \approx 2\pi$, resonant with ±5/2-±1/2.

*Table 1.* Relative oscillator strength for transitions between the $^3H_4$ and $^1D_2$ hyperfine manifolds in $Pr^{3+}$:$Y_2SiO_5$, as calculated from absorption data (c.f. Fig 6). Rows correspond to transitions from different starting states (ground state hyperfine levels). Columns correspond to transitions to different excited state hyperfine levels. All entries have an absolute uncertainty less than ±0.01.

*Table 2*. Rabi frequencies, $\Omega_{ge}$, attained when driving the nine different hyperfine transitions using resonant light with an intensity of $I \approx 250$ W/cm$^2$. The entries are proportional to the transition dipole moments along the electric field polarisation, $\mu$, of the transitions and to the square root of the oscillator strengths (c.f. Table 1). The uncertainty in the absolute values is of the order of ±20%.

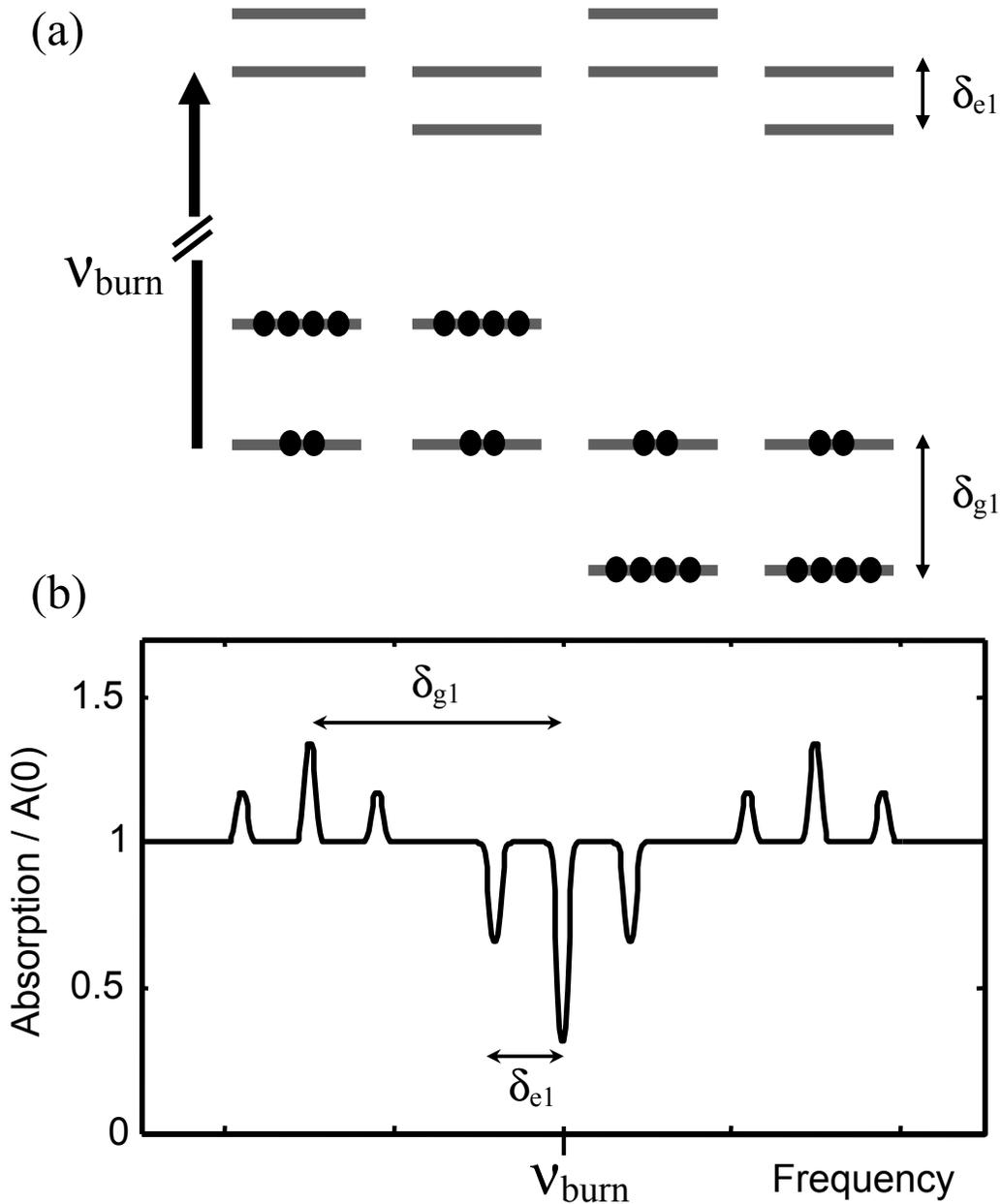

*Figure 1.* Principles of spectral holeburning through optical pumping and redistribution of absorbers between sublevels in the ground state. The example assumes absorbers with two sublevels in both ground and excited states and that the splitting of the ground state is larger than that of the excited state. (a) Due to inhomogeneous broadening, different subsets of absorbers will have transitions between different sublevels shifted into resonance with the applied light. The figure shows the energy levels of four different subsets, all resonant with the burning laser, but on different transitions. (b) Absorption spectrum after burning at frequency $\nu_{burn}$. The position of side-holes and anti-holes directly give the separations between the sublevels.

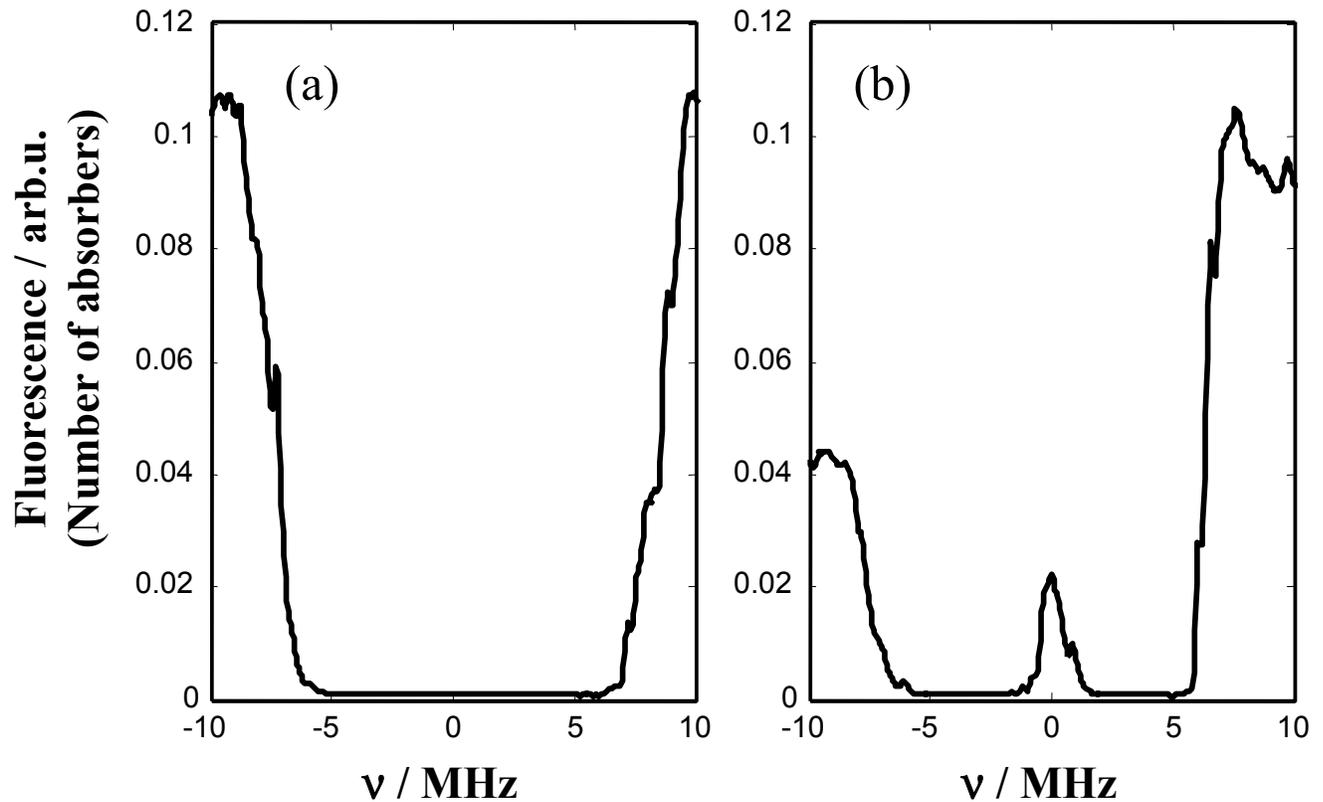

*Figure 2*. Spectral holeburning in the inhomogeneously broadened $^7F_0$ - $^5D_0$ transition in $Eu^{3+}$:$YAlO_3$. The amount of $Eu^{3+}$ ions absorbing at a particular frequency has been monitored by recording laser induced fluorescence. (a) A 12 MHz wide spectral interval has been completely emptied of absorbing ions. At laser frequencies inside the pit, the signal is constant and less than 1% of the fluorescence signal before burning. (b) An ensemble of ions has been moved back into the pit by shifting the laser frequency 69 MHz from the centre of the pit and thus burning on an auxiliary hyperfine level in the ground state. The resolution of the spectral features in this experiment was limited by laser jitter of the order of 2 MHz.

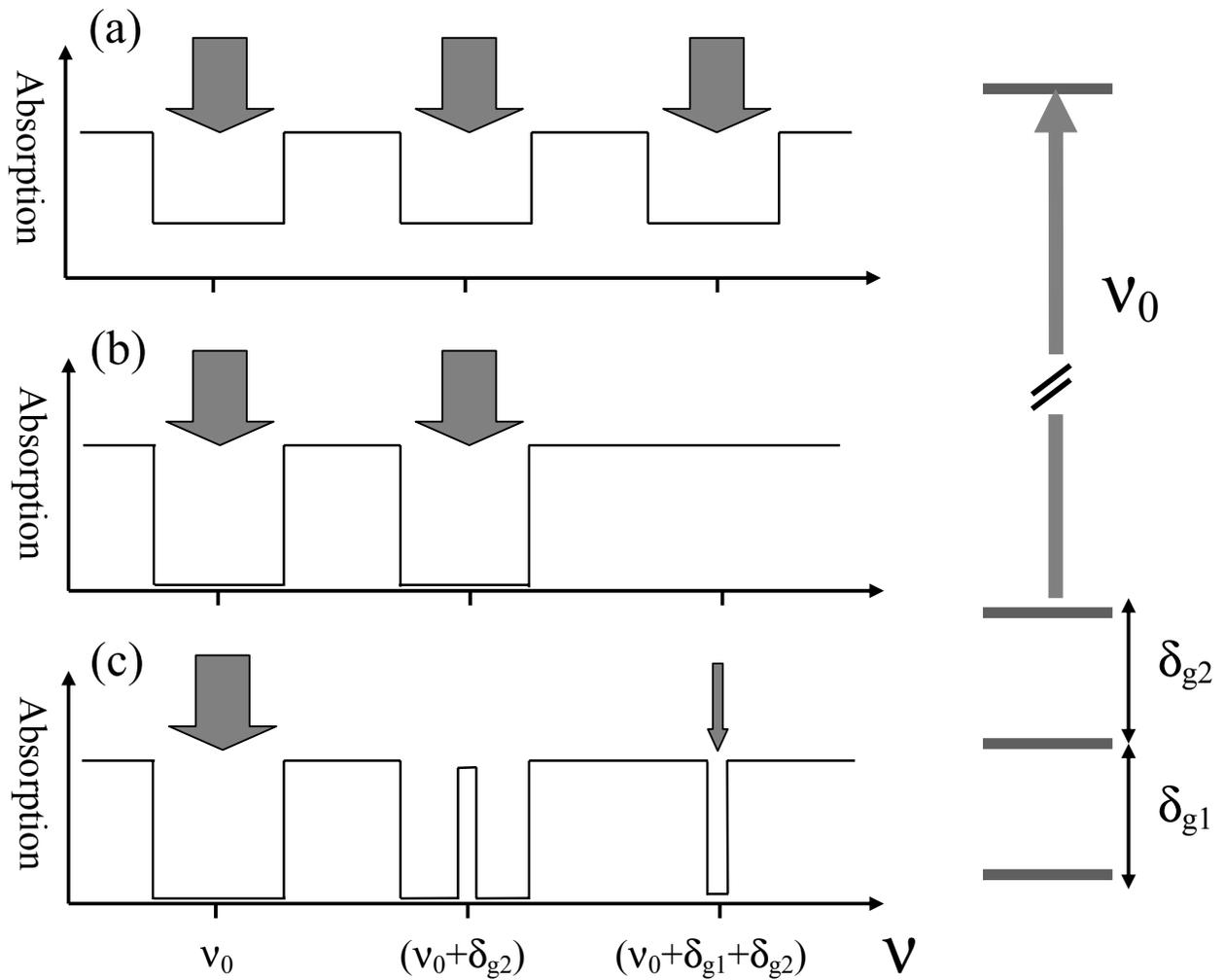

*Figure 3*. *Simplified holeburning sequence, assuming absorbers with three sublevels in the ground state and a single excited state. Energy level diagram is shown for a selected subset of absorbers at a specific position within the inhomogeneous absorption line. (a) Burning at frequencies* $\nu_0$, $(\nu_0 + \delta_2^g)$ *and* $(\nu_0 + \delta_1^g + \delta_2^g)$ *removes absorbers for which these frequencies correspond to other transitions than for the selected subset. (b) Burning at frequencies* $\nu_0$ *and* $(\nu_0 + \delta_2^g)$ *creates two wide spectral pits and collects the selected subset of absorbers in an auxiliary state. (c) Absorbers within a narrow spectral interval at* $(\nu_0 + \delta_1^g + \delta_2^g)$ *are excited from the auxiliary state, from where they relax to form a narrow peak, absorbing from a selected ground state sublevel.*

| | |
|---|---|
| $\lambda\ [^3H_4(1) - {}^1D_2(1)]$ | 605.977 nm |
| $\Gamma_{hom}$ | 970 Hz |
| $\Gamma_{inhom}$ | ~5 GHz |
| $T_1$ | 164 μs |
| $T_2$ | 152 μs |
| $T_{hf}$ (Hole lifetime) | ~100 s |
| $\alpha$ | 20 cm$^{-1}$ |

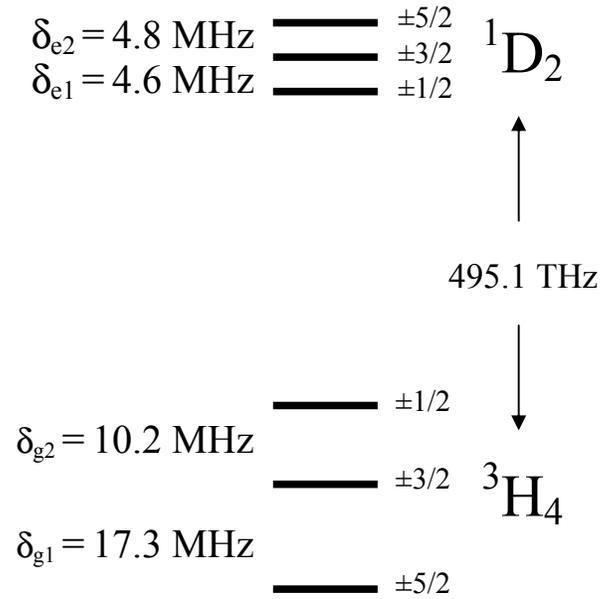

**Figure 4**. *Spectrum and relaxation data for the $^3H_4$ - $^1D_2$ transition of site I Pr$^{3+}$ ions in $Y_2SiO_5$. $T_1$, $T_2$ and $\Gamma_{hom}$ are given according to Ref. 27.*

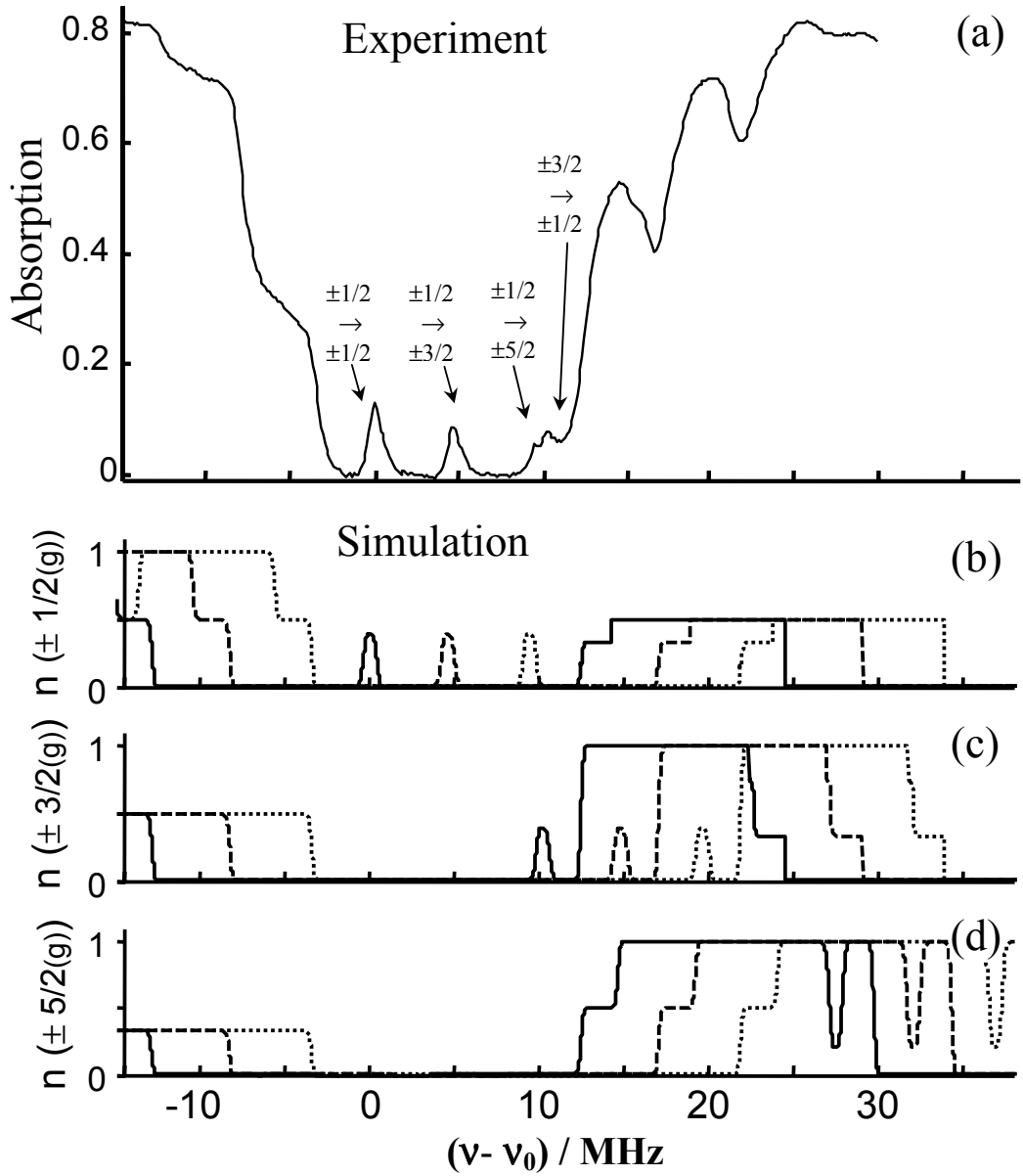

*Figure 5*. (a) The experimental result of applying the holeburning sequence detailed in Sect. B to $Pr^{3+}:Y_2SiO_5$. A group of ions is prepared in state $\pm 1/2$, such that they absorb to each of the excited state hyperfine levels on frequencies inside a previously emptied spectral pit. In this experiment, the spectral resolution was limited by a fast read-out chirp rate ($r \approx 1$ MHz/$\mu$s) and the peak corresponding to the transition $\pm 1/2$-$\pm 5/2$ was not resolved. C.f. Fig. 6a for a close-up of the central region. (b)-(d) Relative number of ions absorbing at each optical frequency from the three ground state hyperfine levels, after applying the sequence of burning pulses in Sect. B. (b) Ions in state $\pm 1/2$, (c) ions in state $\pm 3/2$ and (d) ions in state $\pm 5/2$. Each ion can absorb at three different frequencies. Solid lines: Absorption to the $\pm 1/2$ hyperfine level in the excited state. Dashed lines: Absorption to excited state $\pm 3/2$. Dotted lines: Absorption to excited state $\pm 5/2$.
2222

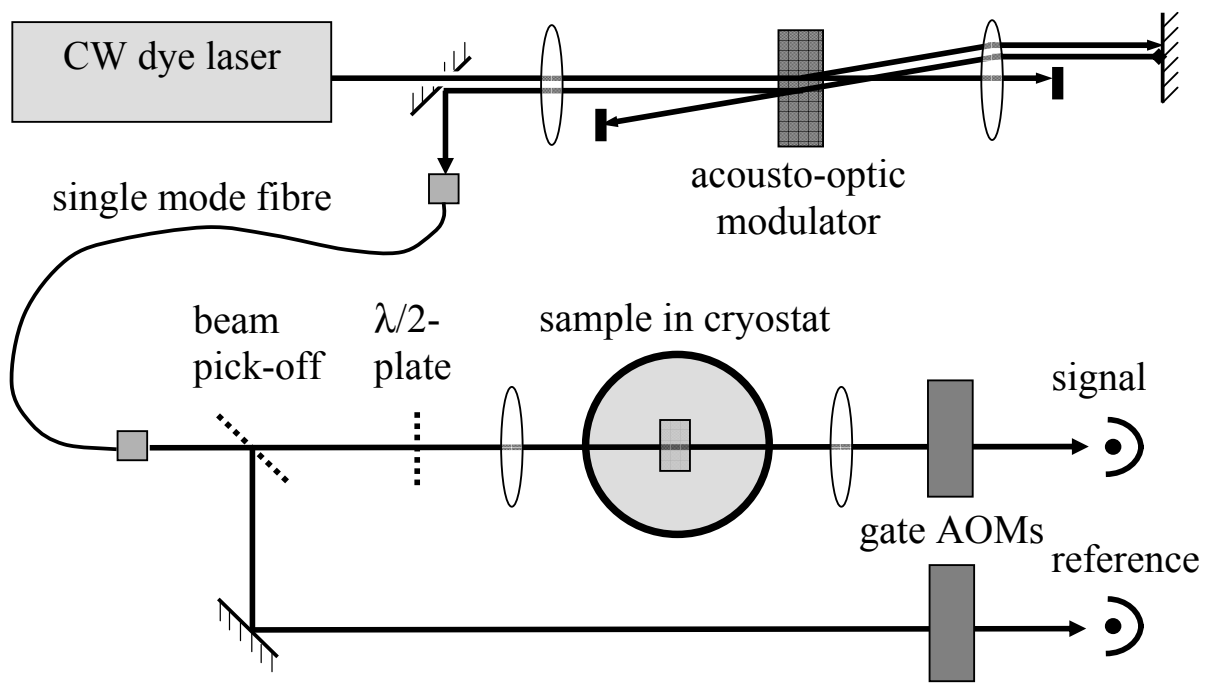

*Figure 6*. *Set-up used in the experiments. See the text, Sect. IV, for details.*

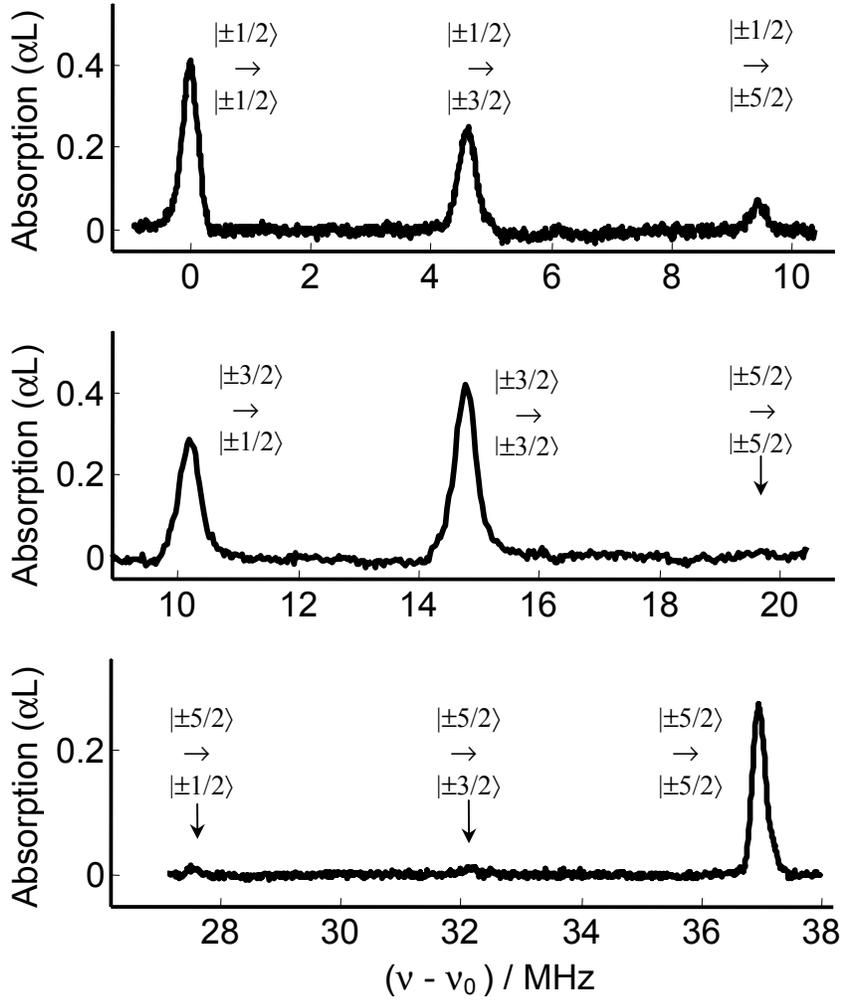

***Figure 7**. Ions absorbing on specific hyperfine transitions, inside the inhomogeneously broadened $^3H_4$ - $^1D_2$ transition in $Pr^{3+}$:$Y_2SiO_5$. The three curves correspond to three separate experiments. The three peaks of varying size, in each curve, are due to the same ensemble of ions absorbing on different transitions. The intensity was attenuated by a factor $10^3$ during read-out, in order not to cause significant excitation when probing the transitions. The scan rate during read-out was of the order of $r \approx 0.1$ MHz/$\mu$s. Single-shot measurements.*

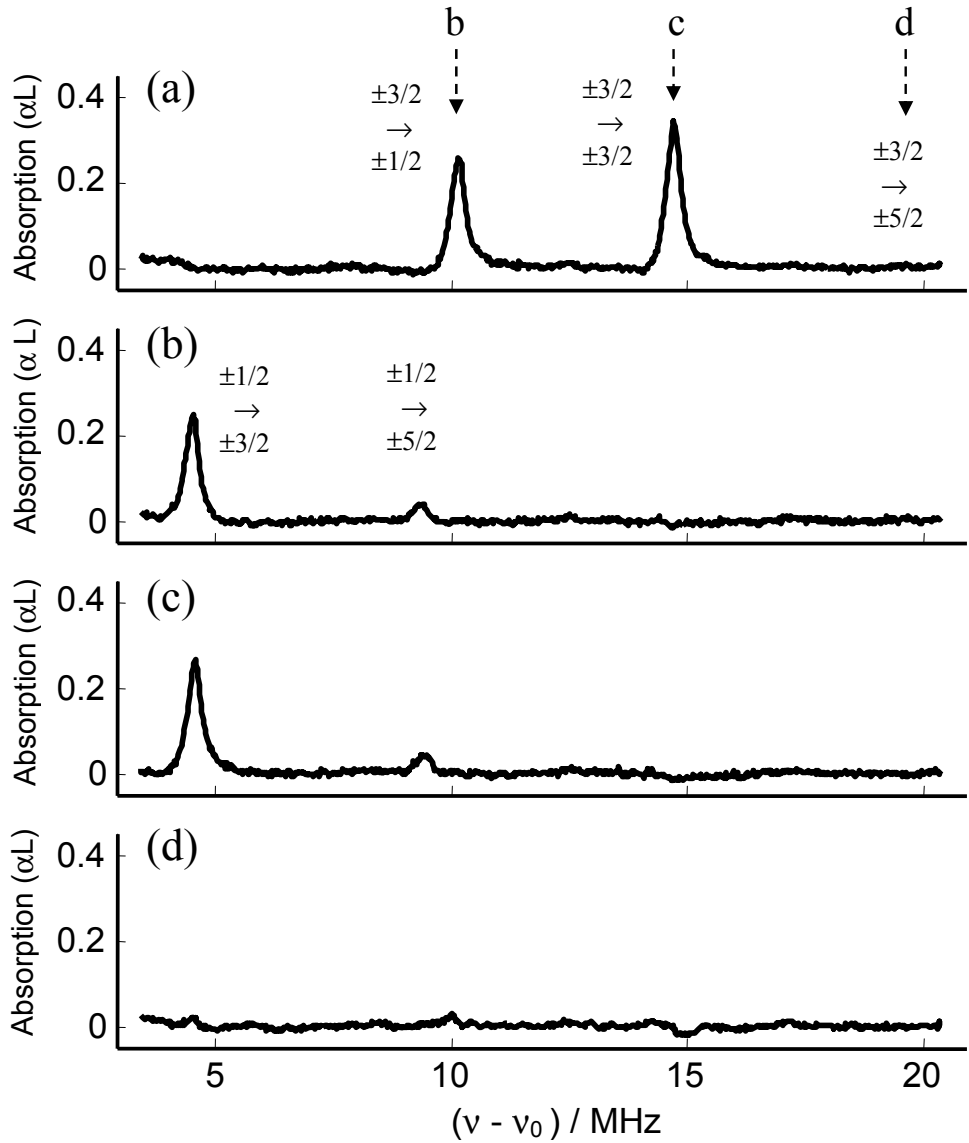

*Figure 8.* *Incoherent burning on transitions going from state ±3/2. (a) A group of ions have been prepared in state ±3/2, inside a wide region emptied of other absorbing ions. Absorption on transition ±3/2-±5/2 is barely detectable. (b)-(d) In three separate experiments, a large number of burning pulses are applied at the frequencies indicated in (a). When the ions are pumped away via (b) the ±1/2 or (c) the ±3/2 hyperfine levels in the excited state, a large fraction of the ions relax to ±1/2 in the ground state and are detected after the burning. (d) When the ions are pumped away via the ±5/2 excited state, they disappear from the monitored region, indicating that they have relaxed to ±5/2 in the ground state, since all transitions from this state are resonant with higher frequencies. Moderate burning at other frequencies inside the pit, i.e. not resonant with any of the peaks, does not change the absorption spectrum. Single-shot measurements.*

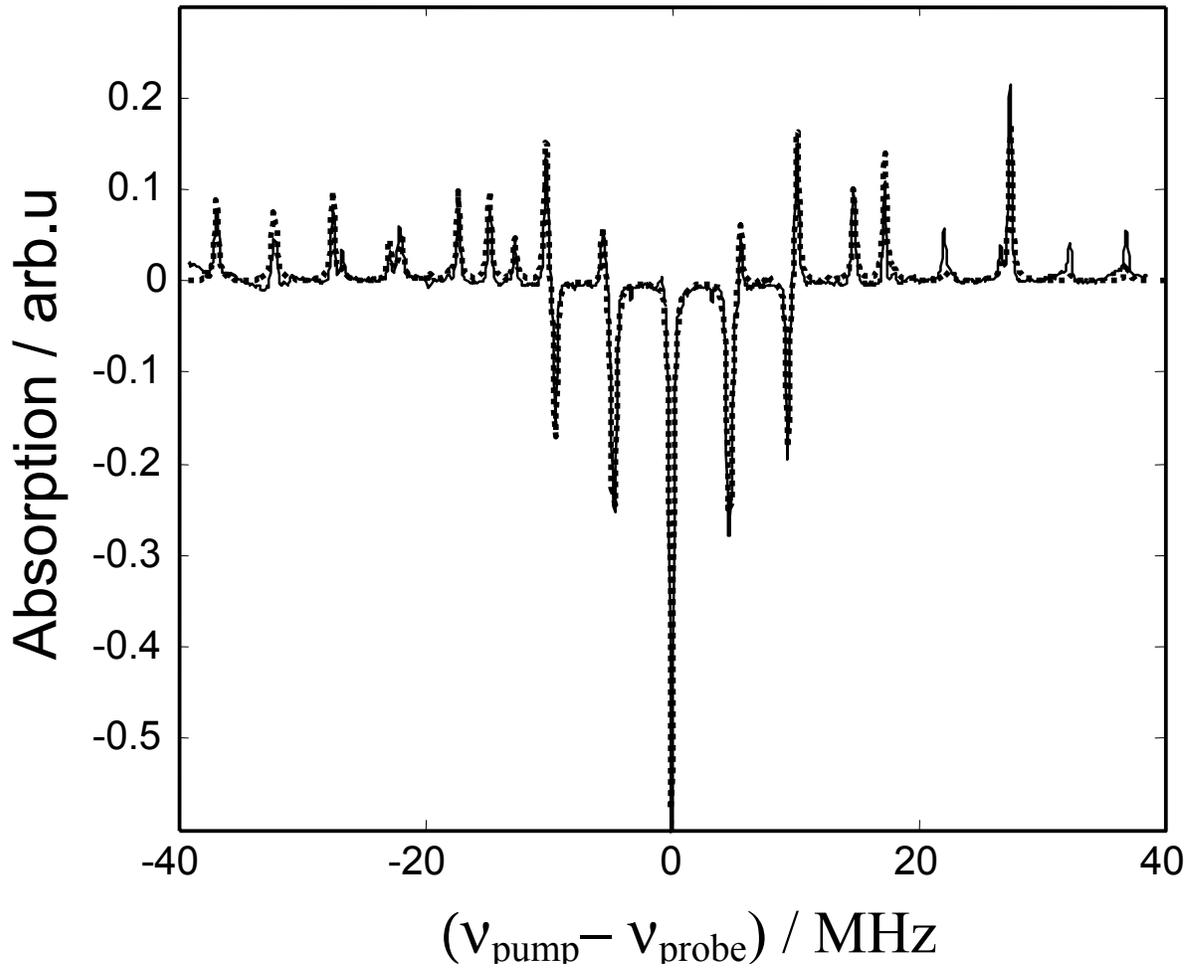

***Figure 9***. *Solid line: Experimental holeburning spectrum in the $^3H_4$ - $^1D_2$ transition in site I $Pr^{3+}$ in $Y_2SiO_5$. The hole was created by applying a burn pulse during 1 s. The data is an average of 30 read-out scans. The spectral resolution is limited by laser jitter during burning and read-out. Dashed line: Simulated holeburning spectrum using hyperfine parameters from Table 1.*

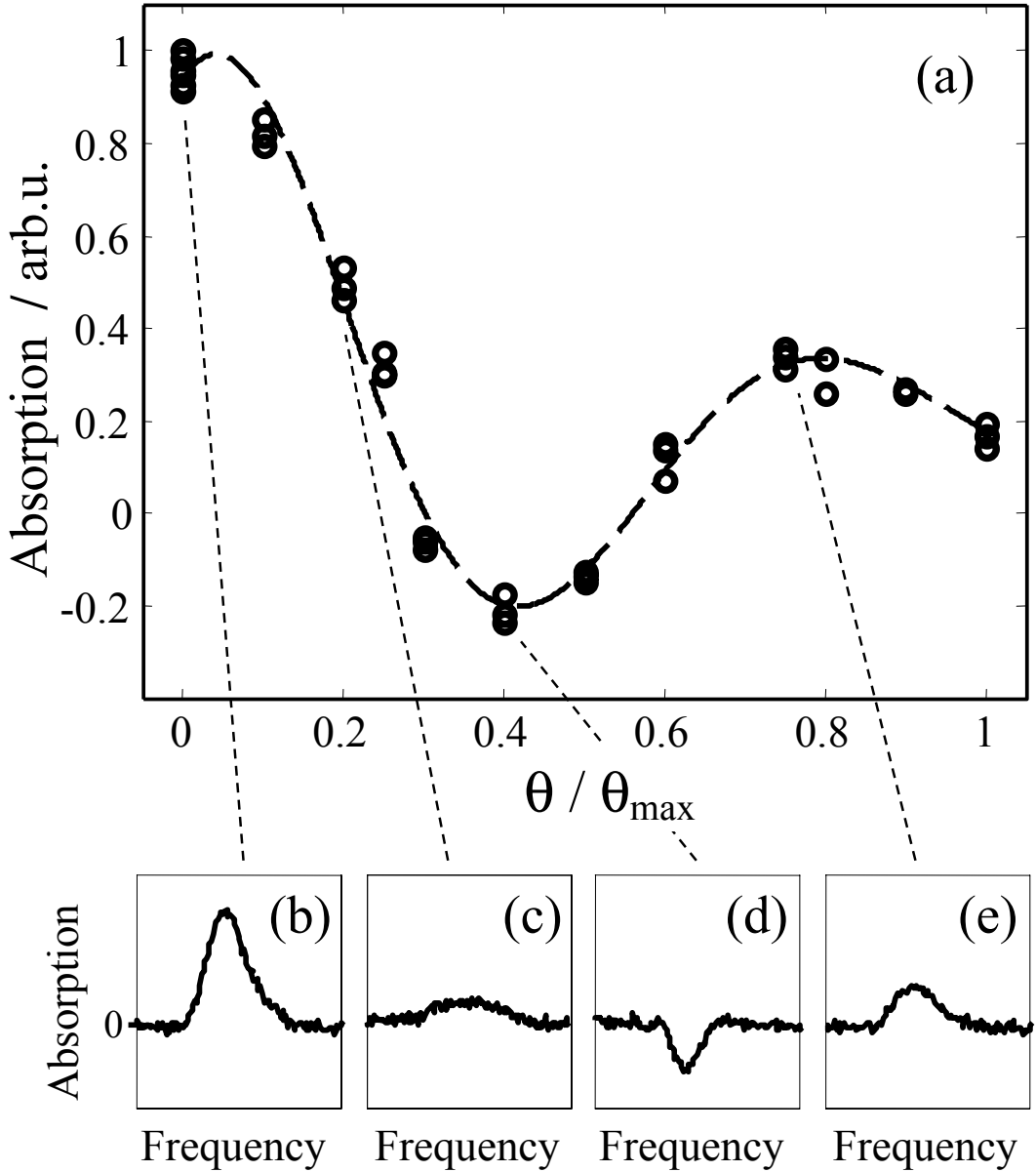

*Figure 10.* Coherent driving of the ±1/2-±3/2 transition. Pulses with a variable pulse area ($\theta = \Omega t$) was applied to the transition using 3 $\mu$s long pulses with varying intensity. The population difference between the ground (±1/2(g)) and optically excited (±3/2(e)) states, after the pulse, was monitored by recording the transmission at the resonance frequency. (b)-(e) show an ensemble of ions absorbing/emitting on the transition after the application of a pulse with (b) $\theta = 0$, (c) $\theta \approx \pi/3$, (d) $\theta \approx \pi$ and (e) $\theta \approx 2\pi$.

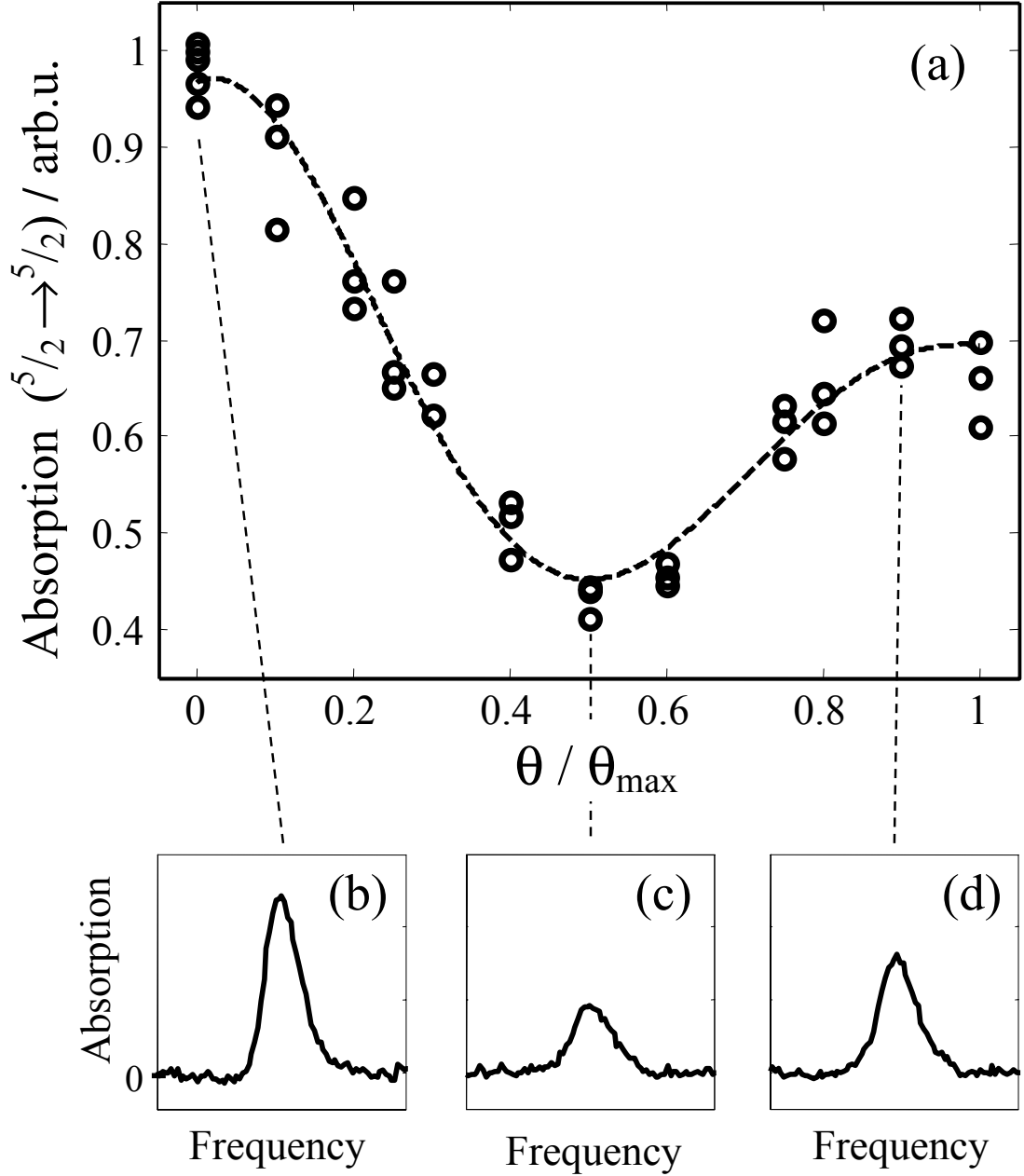

*Figure 11.* Coherent driving of the ±5/2-±1/2 transition. Pulses with a variable pulse area ($\theta = \Omega t$) was applied to the transition using 3 µs long pulses with varying intensity. The number of ions left in the ground state, ±5/2(g), after the pulse, was monitored by recording the absorption on the ±5/2-±5/2 transition. (b)-(d) show an ensemble of ions absorbing on ±5/2-±5/2 after the application of a pulse with (b) $\theta = 0$, (c) $\theta \approx \pi$ and (d) $\theta \approx 2\pi$, resonant with ±5/2-±1/2.

| g \ e | ±1/2 | ±3/2 | ±5/2 |
|---|---|---|---|
| ±1/2 | 0.55 | 0.38 | 0.07 |
| ±3/2 | 0.40 | 0.60 | 0.01 |
| ±5/2 | 0.05 | 0.02 | 0.93 |

***Table 1.** Relative oscillator strength for transitions between the $^3H_4$ and $^1D_2$ hyperfine manifolds in $Pr^{3+}$:$Y_2SiO_5$, as calculated from absorption data (c.f. Fig 6). Rows correspond to transitions from different starting states (ground state hyperfine levels). Columns correspond to transitions to different excited state hyperfine levels. All entries have an absolute uncertainty less than ±0.01.*

$\Omega_{g\leftrightarrow e} / 2\pi \cdot \mathrm{MHz}$

| g \ e | ±1/2 | ±3/2 | ±5/2 |
|---|---|---|---|
| ±1/2 | 1.4 | 1.0 | 0.4 |
| ±3/2 | 1.0 | 1.6 | < 0.2 |
| ±5/2 | 0.4 | 0.3 | 1.8 |

***Table 2**. Rabi frequencies, $\Omega_{ge}$, attained when driving the nine different hyperfine transitions using resonant light with an intensity of $I \approx 250$ W/cm$^2$. The entries are proportional to the transition dipole moments along the electric field polarisation, $\mu$, of the transitions and to the square root of the oscillator strengths (c.f. Table 1). The uncertainty in the absolute values is of the order of ±20%.*